%% file: latex18.tex
\pdfoutput=1

\documentclass{article}
\usepackage{locata,amsmath,graphicx,url,times}

\usepackage{multirow}
\usepackage{amssymb}
\usepackage{subfigure}
\usepackage[british]{babel}
\usepackage[normalem]{ulem}
\usepackage{breqn}
\usepackage{acronym}
\usepackage[numbers,sort&compress,square]{natbib}
\usepackage{url}
\usepackage{color}
\title{LOCATA challenge: Speaker Localization with A Planar Array}

\twoauthors
  {Xinyuan Qian, Andrea Cavallaro}
{Centre for Intelligent Sensing\\
Queen Mary University of London, UK \\
     x.qian@qmul.ac.uk, 
a.cavallaro@qmul.ac.uk}
  {Alessio Brutti, Maurizio Omologo}
    {Fondazione Bruno Kessler\\
Trento, Italy \\
brutti@fbk.eu, 
omologo@fbk.eu}

\def\L{{\cal L}}

\def\pos{{\bf p}} % generic 3D position

\def\state{{\bf s}} % state 
\begin{document}
\include{acronyms}

\ninept
\maketitle

\begin{sloppy}

\begin{abstract}
This document describes our submission to the 2018 \ac{LOCATA} challenge (Tasks 1, 3, 5).  We estimate the 3D position of a speaker using the~\ac{GCF} computed from multiple microphone pairs of a DICIT planar array. One of the main challenges when using such an array with omnidirectional microphones is the front-back ambiguity, which is particularly evident in Task 5. We address this challenge by post-processing the peaks of the \ac{GCF}  and exploiting the attenuation introduced by the frame of the array. Moreover, the intermittent nature of speech and the changing orientation of the speaker make localization difficult. For Tasks 3 and 5, we also employ a~\ac{PF} that favors the spatio-temporal continuity of the localization results. 
\end{abstract}

\begin{keywords}
GCF, Particle Filter, 3D Speaker Tracking, Front-back Ambiguity
\end{keywords}

%%%%%%%%%%%%%%%%%%%%%%%%%%%%%%%%%%%
%\section{Introduction}

\section{Introduction}
\label{sec:intro}
% (INTRODUCTION)
The \ac{LOCATA} corpus includes a development dataset (3 recordings for each task) and a test dataset (13 recordings for static speakers and 5 recordings for moving speakers), with different microphone configurations. The audio signals were recorded at $48$ kHz in an acoustic environment with ambient noise and room reverberation time $T_{60}\approx 0.55 s$ \cite{locata2018heinrich}. For our submission, we considered tasks with one speaker, namely Task 1, Task 3 and Task 5. 

In {Task 1} static microphone arrays recorded a static loudspeaker reproducing a subset of the CSTR VCTK database \cite{Veaux2018english} (newspaper read English sentences). The loudspeaker is placed at different 3D positions and each recording lasts  $2\sim4$ seconds. In {Task 3} static microphone arrays recorded a moving human speaker walking with different head orientations, mostly keeping the mouth at the same height. Each recording lasts $20\sim 30$ seconds. The speaker trajectories and the microphone positions of the 3 recordings of Task 3 are reported in Fig.~\ref{subfig:task3trajectory}. In {Task 5} moving microphone arrays recorded a moving human speaker. Measurement noise is mixed with traffic noise (from outside the recording environment) and with the noise caused by the motion of a moving trolley (which, however, did not appear a critical challenge in our experiments). Each recording lasts $20\sim 50$ seconds. 

Among the 4 microphone arrays available in the challenge, we used the planar array, namely the DICIT array~\cite{locata2018heinrich}. Our goal is to estimate the 3D position, $\pos_{t}$, of the sound source over time $t$. The harmonic spacing between the microphones of the array produces a set of nested sub-arrays with spacings of 4, 8, 16 and 32 cm. The total length is 2.24 m. The array features two microphones located 0.32 m above the outermost microphones to span the vertical coordinate. 
% We do not use those microphones in our system.
Note that the DICIT array is placed in the middle of the room and the speaker moves not only in front of the array (positive $y$-direction in Fig.\ref{subfig:dicitarray}) but also behind the array. 
As mentioned in \cite{Ottoy2016an}, it is not possible to distinguish angles/points from the front or the back of the array with a linear array of omnidirectional microphones. We will exploit the physical support of the array to address this problem.

% In this document, we introduce a method that solves the front-back ambiguity problem, by re-using the intermediate results from the \ac{GCF} localization approach, without adding heavy computational cost.

% \begin{figure}[!bt]
% \begin{center} 
% \subfigure[]{\label{subfig:environment}
% \includegraphics[width=0.45\columnwidth]{figures/DICIT_environment.PNG}}
% \subfigure[]{\label{subfig:dicit}
% \includegraphics[width=0.45\columnwidth]{figures/DICIT_micpositions.PNG}}
% \caption{(a) recoding environment with room dimensions of about $7.1 m\times 9.8 m \times 3 m$; (b) DICIT array: microphone positions and local coordinate system.}
% \label{fig:recordingenv}
% \end{center}
% \end{figure}
\begin{figure}[!bt]
\begin{center} 
\subfigure[]{\label{subfig:task3trajectory}
\includegraphics[width=0.85\columnwidth]{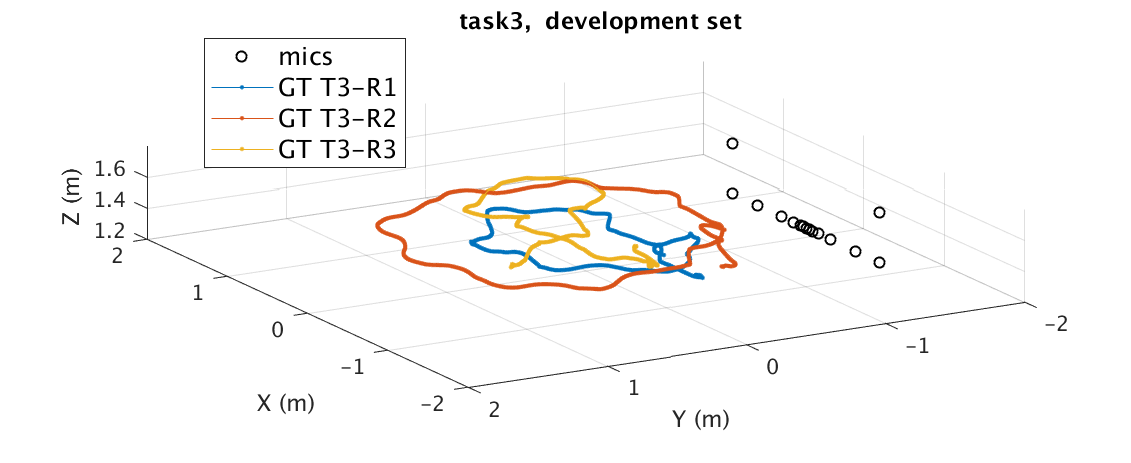}}
\subfigure[]{\label{subfig:dicitarray}
\includegraphics[width=0.8\columnwidth]{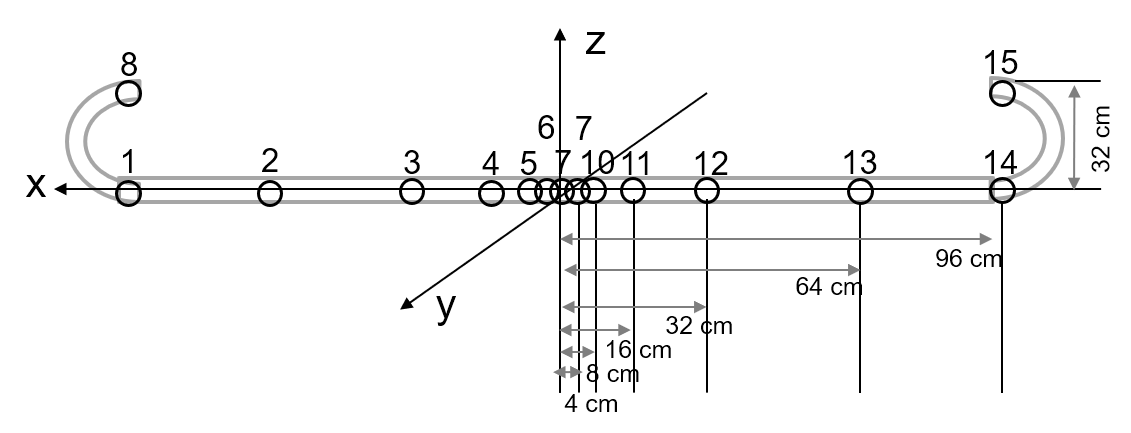}}
\caption{(a) The relative position of the array and the speaker in recording 1, 2, and 3 of Task 3 (each annotated trajectory is color-coded).  (b) The relative positions of the microphones (mics) in the \ac{DICIT} array. }
\label{fig:recordingenv}
\end{center}
\end{figure}

% given the audio signal, ${\bf s}_{k,t}$, captured by the DICIT array, where $k=1,...,15$ is the microphone index.

% Given the array calibration parameters, the 3D estimate $\hat{\pos}_{t}$ in the world coordinates can be transferred to the DICIT local coordinates, with $\pos^\theta_t$ and $\pos^\phi_t$ representing the azimuth and elevation elements, respectively.

%%%%%%%%%%%%%%%%%%%%%%%%%%%%%%
\section{Methods}
\label{sec:aproach}
%\subsection{Definitions}\label{ssec:defs}

\subsection{Localization}\label{ssec:loc}
An acoustic map represents the plausibility of an active sound source existing in a given spatial position.
%The possibility of existing an active sound source at a given spatial position can be represented by an acoustic map. 
Combining information from multiple pairs to generate a global acoustic map (\ac{GCF} \cite{omologo1998spoken}) leads to more robust and accurate localization results than what observed for each individual microphone pair. 

Let $C_{m}(\tau,t)$ be the \ac{GCC-PHAT} \cite{knapp1976generalized, omologo1994acoustic} computed between the microphones of pair $m$ at time $t$, where $\tau$ is the \ac{TDOA}.
% According to \cite{knapp1976generalized, omologo1994acoustic}, the \ac{GCC-PHAT} at the microphone pair $m$ is computed as:
% \begin{equation}\label{eq:GCC-PHAT}
% C_{m}(\tau,t)=\int_{-\infty}^{+\infty}\frac{S_{m_{1}}(t,f)S^{\ast }_{m_{2}}(t,f)}{\left|S_{m_{1}}(t,f)\right |\left | S^{\ast}_{m_{2}}(t,f) \right | }e^{j2\pi f\tau}df,
% \end{equation}
% where $S_{m_{1}}$ and $S_{m_{2}}$ are the \ac{STFT} computed at the two microphones in the $m^{th}$ pair; $\tau$ indicates the intra-microphone time delay ($\ast$ is the complex conjugate). 
The \ac{GCF} value at a generic 3D point $\textbf{p}$ is thus defined as the averaged sum of the \ac{GCC-PHAT}:
\begin{equation}\label{eq:GCF}
g(\pos,t)=\frac{1}{M}\sum_{m=0}^{M-1}C_{m}\left(\tau_{m}(\pos),t\right) ,
\end{equation}
where $M$ is the total number of microphone pairs and $\tau_{m}(\pos)$ is the \ac{TDOA} observed at pair $m$ when a source is in $\pos$. The localization estimate is the peak of the acoustic map:
\begin{equation}\label{eq:3Dloc}
\hat{\pos}_t=\arg \max_{\pos \in {\bf P}} \ g(\pos,t) ,
\end{equation}
where ${\bf P}$ is a spatial grid with the potential sound source positions. %The \ac{GCF} peak value at time $t$ is computed as:
%\begin{equation}
%g^\prime(t)=g(\hat{\pos}_t,t)
%\end{equation}
 In Task 1 (static speaker), the resulting 3D position is the most frequent estimate over the recording:
\begin{equation}\label{eq:3Dlocstatic}
\hat{\pos}=mode\{\hat{\pos}_1,...,\hat{\pos}_T\} .
\end{equation}
where $mode$ selects the most frequent element of a set and $T$ is the total number of frames.

%%%%%%%%%%%%%%%%%%%%%%%%%%%%%%%%%%%%
\subsection{Tracking}
\label{ssec:tracking}

In Task 3 and 5, we adopt a \ac{PF}~\cite{arulampalam2002tutorial} to track the moving speaker in 3D. Let particle $n$, $n=1,...,N$, at time $t$ be defined in 3D as:
\begin{equation}\label{eq:state}
\state^{(n)}_t=[x_t,y_t,z_t]^\intercal,
\end{equation}
where $\intercal$ indicates transpose. Particles are propagated as:
\begin{equation}\label{eq:propagate}
\state^{(n)}_t=\state^{(n)}_{t-1}+\mathcal{N}(0,\Sigma)  ,
\end{equation}
where $\mathcal{N}(0,\Sigma)$ is Gaussian noise with zero mean and covariance matrix $\Sigma$, whose $z$-element is smaller than the counterparts for $x$ and $y$ as the height of the mouth undergoes smaller changes than the $x$-$y$ position of the speaker.

Instead of using a voice activity detector to find speech segments, we use the absolute values of the \ac{GCF} peak (Eq.~\ref{eq:3Dloc}) as the indicator of the presence of speech and as the measure of the reliability of the observation~\cite{qian20173d}. Thus, the likelihood function to update the particles is: 
\begin{equation}\label{eq:likelihood}
\L^{(n)}_t=
\left\{\begin{matrix}
 \mathcal{N}(\state^{(n)}_t|\hat{\pos}_t,\sigma^2)& \text{if} \ g^\prime(t)\geq  \alpha \gamma\\ 
 g(\state^{(n)}_t,t) & \text{else if}\ g^\prime(t)\geq  \beta \gamma \\ 
 \mathcal{N}(0,1) & otherwise ,
\end{matrix}\right.
\end{equation}
where $\sigma$ is the standard deviation of the Gaussian distribution. $g^\prime(t)=\max_{\pos \in {\bf P}} \ g(\pos,t)$, $\gamma=\max \ \{g^\prime(t)\}^T_{t=1}$ (the maximum \ac{GCF} value over the whole recording, $\alpha$ and $\beta$ are parameters present in percentage to select between high confident and less confident \ac{GCF} estimates. This selective likelihood is designed for quick convergence at the high confident estimates and slow convergence during less confident estimations.

Finally, the speaker's 3D position at time $t$ is estimated as:
\begin{equation}\label{eq:posupdate}
\hat{\pos}_t=\frac{\sum^{N}_{n=1} \L^{(n)}_t \state^{(n)}_t}{\sum^{N}_{n=1} \L^{(n)}_t} \ . 
\end{equation}
% where $N$ indicates the particle number.

In the re-sampling stage, we use \ac{SIR}: particles with higher weights are duplicated while those with lower weights are eliminated.
%%%%%%%%%%%%%%%%%%%%%%%%%%%%%%%%%%%%%

\begin{figure}[!bt]
\begin{center} 
\subfigure[]{\label{subfig:tracj5}
\includegraphics[width=0.7\columnwidth]{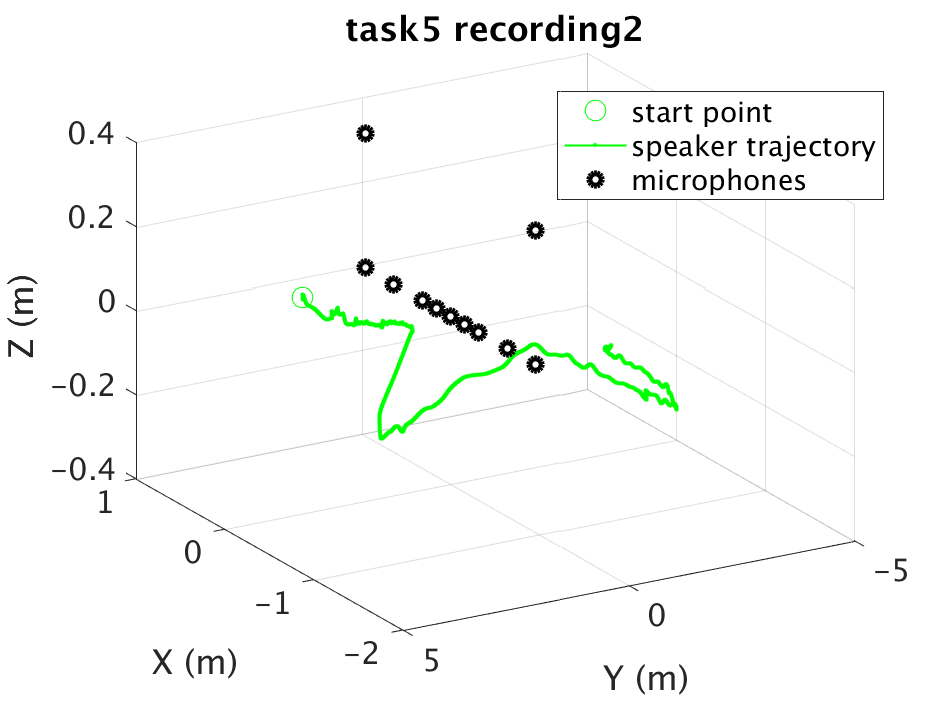}}
%\subfigure[]{\label{subfig:environment}
%\includegraphics[width=0.7\columnwidth]{figures/task5recording2_development2_azel.png}}
\subfigure[]{\label{subfig:gccmax}
\includegraphics[width=0.7\columnwidth]{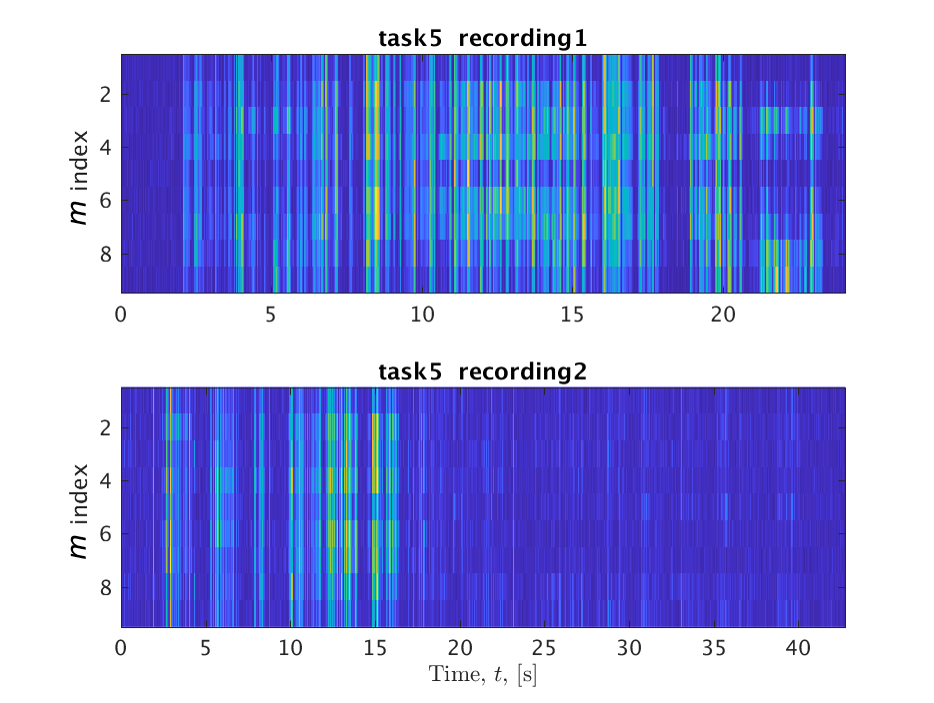}}
\caption{(a) The annotated trajectory of the speaker in the \ac{DICIT} array coordinate system for recording 2, development dataset, Task 5. (b) From top to bottom: the \ac{GCC-PHAT} peaks (yellow) of recording 1 and 2 for different microphone pairs (vertical axis).}
\label{fig:task5trajectory}
\end{center}
\end{figure}

\subsection{Front-back ambiguity}\label{ssec:frontback}

An important issue in Task 5 when the speaker moves to the back of the DICIT array (e.g. in recording 2 of the development set). Fig.~\ref{fig:task5trajectory}(a) shows the trajectory of the moving speaker, marked as green, with a circle indicating the starting point. Microphone positions are plotted in black. The figure illustrates that the speaker (or the array) moves from the front, walks underneath the array, and finally goes to the back, which cannot be distinguished from the front case when using a planar array of omnidirectional microphones. To solve this front-back ambiguity, we propose a novel approach which, without heavy computations, detects when the speaker is behind the array by using the \ac{GCF} peak value and relying on the attenuation due to the array box.

% The azimuth ground truth of the speaker is plotted on the top row of Fig.\ref{fig:task5trajectory} (b), where at around $18$ seconds (azimuth$\approx -90^\circ$), the speaker turns to the back. The second row of Fig.\ref{fig:task5trajectory} (b) illustrates the conducted \ac{GCC-PHAT} at the middle microphone pair ($m_1=4,m_2=11$), with the ground truth marked as yellow. We can see that after the speaker turning ($t>18$ $s$), the \ac{GCC-PHAT} ground truth is almost the same as the previous frames, even under varying azimuths. 

% Therefore, the linear planar \ac{DICIT} array cannot distinguish the front and the back, if only applying the \ac{GCC-PHAT} technique. 
According to \cite{brutti2010WOZ}, microphones on the DICIT array are omnidirectional, but they are configured along a metallic frame box and pointed to the front for the purpose of recording sound coming from the forward direction. In this case, when the source is located at back, the direct-path signal is attenuated by the array frame. This results, in practice, in a semi omnidirectional polar pattern, which can be leveraged to solve the front-back issue. Note however that attenuation due to the array frame will critically affect the localization performance when the source is behind the array.

Fig.~\ref{subfig:gccmax} shows the \ac{GCC-PHAT} peak value of recording 1 and 2, of Task 5. The horizontal axis indicates the time and the vertical axis shows the microphone pair index. Yellow and blue correspond to higher and lower values, respectively. The speaker goes from the front to the back in the second half of recording 2, while always staying in front of the array in recording 1. Therefore, the \ac{GCC-PHAT} peak value in recording 2 is degraded during the last periods. The figure shows variations of the \ac{GCC-PHAT} peak value among different pairs, at the same time in the same recording, which may be related to the speaker's relative position to the pairs. Considering different \ac{GCC-PHAT} performance among pairs at the same time, instead of using individual \ac{GCC-PHAT}, we use the \ac{GCF} peak value to solve the front-back ambiguity. However, because the \ac{GCF} peak value includes various oscillations from frame to frame (see top rows of Fig.~\ref{fig:GCFratio}), which may result from instant silence and cannot be used directly, we propose to average the \ac{GCF} peak value, in the forth and back way, and use the ratio to find the speaker turning time. 

Let $\bar{g}^+(t)$ and $\bar{g}^-(t)$ be the forward and backward \ac{GCF} peak averages, at time index $t$:
\begin{eqnarray}\label{eq:GCFforwardbackward}
\bar{g}^+(t)&=&\frac{1}{t}\sum^t_{1}g^\prime(t) \ ,\\
\bar{g}^-(t)&=&\frac{1}{T-t+1}\sum^T_{t}g^\prime(t) \ ,
\end{eqnarray}

We introduce the front-back \ac{GCF} ratio as:
\begin{equation}\label{eq:GCFforback}
g^{\pm}(t)= \frac{1}{g^{\mp}(t)}=\frac{\bar{g}^+(t)}{\bar{g}^-(t)} \ ,
\end{equation}
where $g^{\mp}(t)$ is its reciprocal, indicating the back-front motion. Given the assumption that, in each interval $t=1,\dots,T$, the front-back reversal happens once, the swap frame can be found by looking for the most variant \ac{GCF} ratio:
\begin{equation}\label{eq:frontback}
t^+=\arg \max_{t \in ({t_0},T-{t_0}) } g^{\pm}(t) \ ,
\end{equation}
where $({t_0},T-{t_0})$ is the period during which the turning could happen. Similarly, the back-front candidate frame is defined as:
\begin{equation}\label{eq:backfront}
t^-=\arg \max_{t \in ({t_0},T-{t_0}) } g^{\mp}(t) \ ,
\end{equation}
Finally, the turning frame is estimated as:
\begin{equation}\label{eq:turningpoint} 
t^\prime=
\left\{\begin{matrix}
 t^+ & g^{\pm}(t)\geq  \kappa,\ g^{\pm}(t)> g^{\mp}(t) \\ 
 t^- & g^{\mp}(t)\geq  \kappa,\ g^{\mp}(t)> g^{\pm}(t) \\ 
\o & otherwise \ .
\end{matrix}\right. 
\end{equation}
where $\kappa$ is a threshold deriving from the development set analysis.

\begin{figure}[!tb]
\begin{center} 
\subfigure[]{\label{subfig:T5R2_frontback}
\includegraphics[width=0.85\columnwidth]{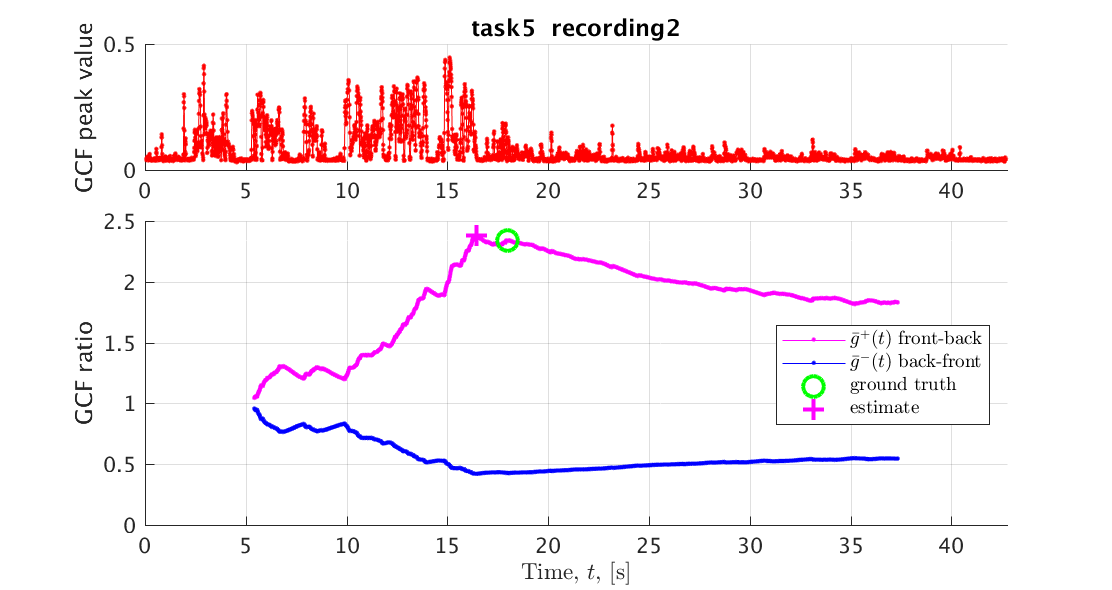}}
\subfigure[]{\label{subfig:environment}
\includegraphics[width=0.85\columnwidth]{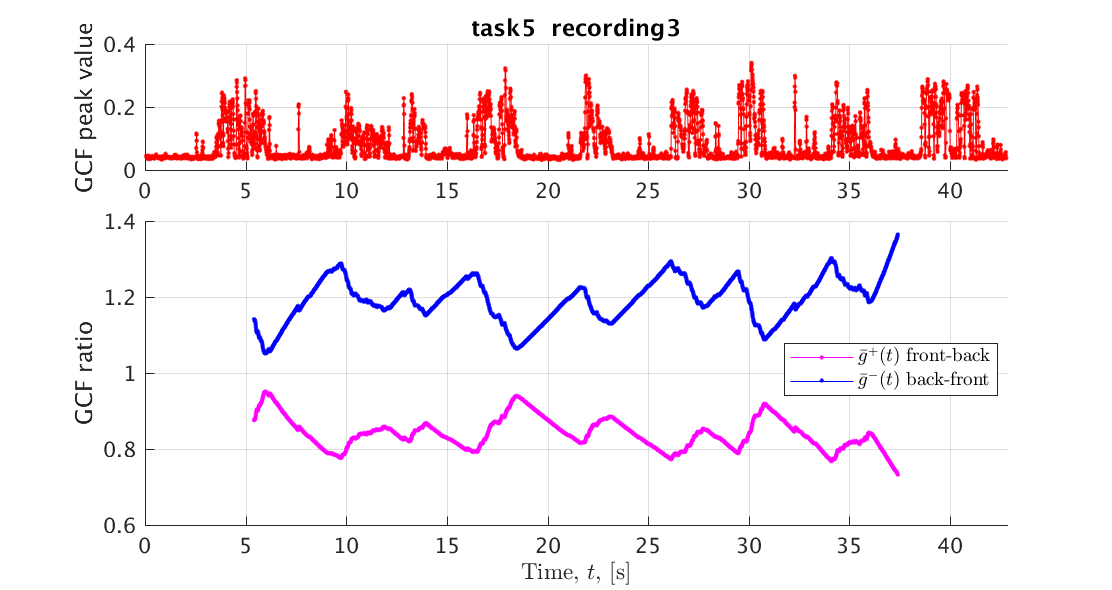}}
\caption{Illustration of the \ac{GCF} peak value, the front-back ($g^{\pm}(t)$, magenta) and the back-front ($g^{\mp}(t)$, blue) ratio of Task 5, development set: (a) the speaker walks behind the array at $t\approx 18$ $s$ in recording 2; (b) the speaker always moves in the front of the array in recording 3.}
\label{fig:GCFratio}
\end{center}
\end{figure}

The top row of Fig.~\ref{fig:GCFratio}(a) and (b) displays the \ac{GCF} peak value over frames of recording 2 and 3 of Task 5, respectively. The bottom row illustrates the front-back and back-front \ac{GCF} ratio. In Fig.~\ref{subfig:T5R2_frontback}, the speaker moves from the front to the back, with $g^{\pm}(t)$ exceeding the threshold ($\kappa$) and reaching its maximum value at around 16.4 seconds (magenta cross), which is very close to the turning point (green circle). In (b), the speaker always stays in front of the array, so both $g^{\pm}(t)$ and $g^{\mp}(t)$ don't have an obvious global peak. Instead, their value oscillates around a constant, which results from changes between speech and non-speech segments.

Once the turning point $t^\prime$ is found, the position estimates are reversed along the $y$-direction (see \cite{locata2018heinrich} for the array's local coordinates information) to achieve the corrected position estimates $\hat{\pos}_{t}$.

%%%%%%%%%%%%%%%%%%%%%%%%%%%%%%%%%%%%%%
\subsection{Outlier smoothing}
\label{ssec:outliersmooth}

Because no voice activity detector is used, the \ac{GCF} cannot distinguish a sound produced by the human target from that produced by a noise source. Thus, the \ac{PF}  may be misled by consecutive false positives (continuous noise source). Using a small particle velocity could help to avoid this problem, but will make the tracker prone to local maxima. Therefore, assuming that the velocity of the speaker is unlikely to change abruptly, when the estimated instant velocity at time $t$ is larger than a threshold $\upsilon$, we iteratively remove 3D tracking results in a short interval (e.g.~$[t-1,t+2]$) and replace them by interpolating nearby estimations. This smoothing process stops when the maximum velocity is smaller than $\upsilon$ or after a maximum number of iterations.

%%%%%%%%%%%%%%%%%%%%%%%%%%%%%%%
\section{Experiments}
\label{sec:experiments}

%\subsection{Performance measure}\label{ssec:metrics}

\subsection{Parameters}

The parameters were chosen on the development recordings, using the same values for the three tasks, except the window length.

{\bf{Window lengths}}. The \ac{STFT} window length plays a significant role in localization based on \ac{GCC-PHAT}, due to the assumption of spatial stationarity of the sound source inside the windowed audio sequences.
% the spatial properties of the speaker are time-invariant in each window segment. 
% Because speakers are moving freely during recordings, their velocities vary from frame to frame. 
% The slower a speaker moves, the smaller the variation of his acoustic characteristics.
For a static speaker, a long window length is preferable since the source is always stationary.
When a sound source is moving, the stationary assumption is approximate to be held by using a short window length, which however leads to a more noisy estimation, and to a lower \ac{TDOA} resolution.
% The window length affects the localization results: 
% a longer window might include more silent and noisy audio signals, which may degrade the localization performance. 
Fig.~\ref{fig:FFTer} shows the localization accuracy of the azimuth estimates, when increasing the \ac{STFT} window length from $2^{10}$ to $2^{15}$ points, with the color changing from dark blue to bright yellow. The results show that a longer window leads (mostly) to a better accuracy. We use $2^{14}$ for a static loudspeaker and $2^{12}$ for a moving speaker. Because even the localization error is larger, a shorter window could capture more local variants, where the instant outliers will be handled by the \ac{PF}. Besides, due to the fact that only three recordings are provided in the development set, it is risky to assume that in the test set the speaker is moving with similar slow velocities. 

\begin{figure}[!tb]
\begin{center} 
%\subfigure[]{\label{subfig:environment}
%\includegraphics[width=0.9\columnwidth]{figures/FFTazimuth_error}}
\includegraphics[width=\columnwidth]{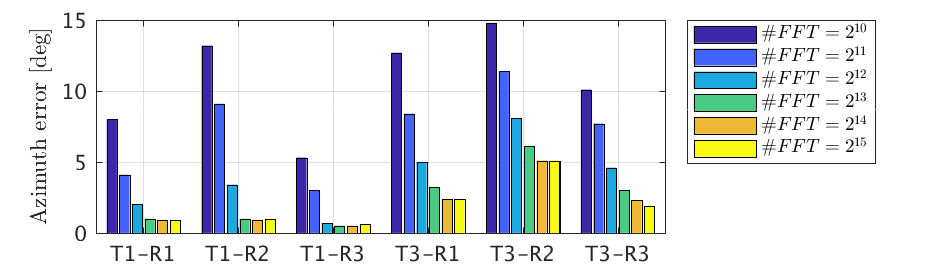}
% \subfigure[]{\label{subfig:environment}
% \includegraphics[width=0.48\columnwidth]{figures/FFT3d_error}}
\caption{Effect of the window length on the localization accuracy of the azimuth estimates (T1-R1 indicates Task 1, recording 1).}
\label{fig:FFTer}
\end{center}
\end{figure}

{\bf{Array pairs}}.
The DICIT array consists of 15 microphones that lead to 105 possible microphone pairs. However, not all the pairs are desirable. For example, the most distant microphones are 1.92 m apart, and thus their audio signals have limited coherence.
% Table \ref{tab:micarraypairs} shows the localization accuracy on Task 1 (static loudspeaker) in azimuth ($\theta$), elevation ($\phi$) and 3D, using different microphone combinations. 
According to our experiments in the development dataset, using the 32 cm pairs gives the best results. As illustrated in \cite{locata2018heinrich}, the 4 cm distance pairs are in the middle of the array where the maximum intra-microphone distance is only 8 cm. In this case, all the sensors experience the similar acoustic environment, which is more sensitive to the local noise. Besides, since our goal is to find the 3D source location, the sub-array with 4 cm spacing is too small to get benefits of the sensor geometric information and it is difficult to reduce the localization uncertainties in 3D. The same reason is applicable to the 8 cm and 16 cm pairs.

{\bf{Other parameters}}.
We define a regular 3D grid with points spaced at 2 cm distance, in the range of $x\in [-3,3]$ m, $y \in \left[-0.1, 4 \right]$ m and $z \in [1.3,1.75]$ m. We use the Blackman window to segment the audio signals for the \ac{DFT} computation. In PF, we use $N=100$ particles. The prediction covariance in Eq.~\ref{eq:propagate} is $\Sigma=diag(0.1,0.1,0.005)$ m. The percentage indices are $\alpha=0.2$ and $\beta=0.1$.
The standard deviation in the likelihood update process is related to the \ac{GCF} localization accuracy using the DICIT array and is set to $\sigma=0.2$  m. The front-back threshold is $\kappa=1.9$.  In the smoothing process, $\upsilon=2$ m/s and the maximum  number of iterations is 15. Finally, we ignore  $650$ frames at the beginning and at the end of each recording.

%%%%%%%%%%%%%%%%%%%%%%%%%%%%%%%%%%%%%%%%%%%%%%%%%%%%%%%%%%%%
\subsection{Comparison with the baseline and discussion}

The performance is evaluated only during the source activity periods using the \ac{MAE}:

 \begin{equation}\label{eq:metrics}
 \varepsilon=\sum_{t \in \tilde{T}} ||\dot{\pos}_t-\hat{\pos}_t||_2 \ .
 \end{equation}
 
where $\dot{\pos}_t$ is the 3D ground truth, $\tilde{T}$ is the set of voice-active frames. This equation is applicable to the \ac{MAE} in azimuth and elevation, by replacing $\dot{\pos}_t$ with $\dot
{\pos}^{\theta}_t$ and $\dot{\pos}^{\phi}_t$, $\hat{\pos}_t$ with $\hat{\pos}^{\theta}_t$ and $\hat{\pos}^{\phi}_t$, respectively. $\dot{\pos}^\theta_t$ and $\dot{\pos}^\phi_t$ are the azimuth and elevation elements transferred from $\dot{\pos}_{t}$ to the DICIT local coordinates, given the array calibration parameters. The same definition is used for $\hat{\pos}^{\theta}_t$ and $\hat{\pos}^{\phi}_t$.

% Given the array calibration parameters, the 3D estimate $\hat{\pos}_{t}$ in the world coordinates can be transferred to the DICIT local coordinates, with $\pos^\theta_t$ and $\pos^\phi_t$ representing the azimuth and elevation elements, respectively.

Table \ref{tab:res} compares our localization and tracking results (average over 5 iterations) with those of the baseline described in \cite{locata2018heinrich}, which uses \ac{MUSIC} \cite{trees2002optimum,dmochowski2007broadband} to provide the \ac{DoA} estimates. %Our method employs tracking in Task 3 and 5, thus leading to a substantial improvement of the localization accuracy. 
The large localization error of Task 5 is caused by the front-back ambiguities, as discussed in Sec. \ref{ssec:frontback}. 
%Apart from this, sparse azimuth resolution and unreasonable microphone pair selection are other reasons lead to the unsatisfactory results. 

Our method outperforms the baseline because of the microphone pair selection; a smaller step-size between consecutive \ac{STFT} blocks, computed with reasonable window length; and the prediction and update process at the required timestamps. 

{\bf Microphone pair selection}. The \ac{MUSIC} baseline only uses the middle three microphones for \ac{DoA} estimation, with a spacing of only 4 cm. In addition, the resolution of the azimuth search space is set to $5^\circ$, which is rather large for an accurate estimate.

{\bf Smaller step-size between consecutive \ac{STFT} blocks, computed with an adequate window length}. The baseline estimates are derived from a $2^{10}$-point \ac{STFT} and are then interpolated to fit the required timestamps. Since the speaker moves slowly, our \ac{STFT} is computed with longer windows leading to the better accuracy (see Fig.~\ref{fig:FFTer}). Besides, our approach estimates the source position at the required timestamps, catching more variations of the adjacent acoustic characteristics,
%{\color{blue}catching more local acoustic variants WHAT DOES THIS MEAN?}, 
without interpolation. 

{\bf Prediction and update process at the required timestamps}. The un-interpolated \ac{MUSIC} estimates are used to provide a smoothed azimuth trajectory using \ac{KF} at the required high-resolution timestamps. Because \ac{MUSIC} is computed at a much lower time resolution, the same observations are used for several~\ac{KF} estimates. As a result, outlier observations in those consecutive frames cannot be filtered out.

\begin{table}[!tb]
\caption{Comparison of the localization (loc) and tracking (trk) results (\ac{MAE}) of our approach and the baseline method on Task (T) 1, 3, and 5.}
\label{tab:res}
\begin{tabular}{c|c|c|c|c|c|c|c|c|}
\cline{2-9}
                                      & \multicolumn{2}{c|}{\textbf{Baseline}} & \multicolumn{6}{c|}{\textbf{Our method}}                                                                                 \\ \cline{2-9}
  & \multicolumn{2}{c|}{$\theta \ (^\circ)$}  & \multicolumn{2}{c|}{$ \theta \ (^\circ)$} & \multicolumn{2}{c|}{$ \phi \ (^\circ)$} & \multicolumn{2}{c|}{3D (m)} \\ \cline{2-9}
 & loc    & trk    & loc     & trk     & loc      & trk     & loc   & trk  \\ \hline \hline
\multicolumn{1}{|c|}{T1} & 50.0              & 16.0               & 0.8              & -                  & 4.1               & -                   & 0.19           & -               \\ %\hline
\multicolumn{1}{|c|}{T3} & 70.9              & 36.6               & 6.0              & 1.8                & 2.9              & 2.0                 & 0.41           & 0.15           \\ %\hline
\multicolumn{1}{|c|}{T5} & 81.0              & 25.7               & 20.5             & 2.7                & 3.3             & 2.4                &1.08           & 0.18           \\ \hline \hline
\end{tabular}
\end{table}

\subsection{Limitations}
%The DICIT array in this \ac{LOCATA} corpus is embedded on a thinner and lighter frame, compared to the one configured in \cite{brutti2008localization}.
%Therefore, we presume that when the sound source is located at the array's back, the direct-path signal has less attenuation than the one used in \cite{brutti2008localization}, which makes the front-back distinction more difficult. 

%%%%%%% limitations discussion %%%%%%%%%%% 

There are two main limitations of the proposed approach, namely its computational cost and the reliability of its likelihood function under multiple sound sources. 

A limitation of our approach is its computational cost, which is proportional to the number of grid points (see Table~\ref{tab:gridstep}). For example, when the grid resolution is reduced from 0.5 m to 0.01 m, the elapsed time, which includes the process of 3D grid creation, \ac{GCC-PHAT} computation and \ac{GCF} retrieval, increases exponentially, while the localization error becomes stable below 0.05 m. Since 0.02 m and 0.01 m have similar localization errors (i.e.~the difference could be attributed to inaccuracies in the ground-truth annotation), we chose 0.02m as grid resolution. When grid points are few, the \ac{GCC-PHAT} computation dominates the elapsed time. Since $2^{14}$ and $2^{12}$ window lengths are used in T1 and T3 respectively, the elapsed time in T1 is longer than in T3 in the first columns.

The reliability of the likelihood function in Eq.~\ref{eq:likelihood} decreases with additional sound sources. For example, a second sound source is active in recording 2 of Task 3 and generates a high \ac{GCF} peak: Fig.~\ref{fig:centernoise} shows that black dots representing particles converge toward this distractor that is far from the ground truth (green cross). In this case, because of the large value of the corresponding \ac{GCF} peak value, the localization accuracy degrades.

\begin{figure}[!tb]
\begin{center} 
\includegraphics[width=0.85\columnwidth]{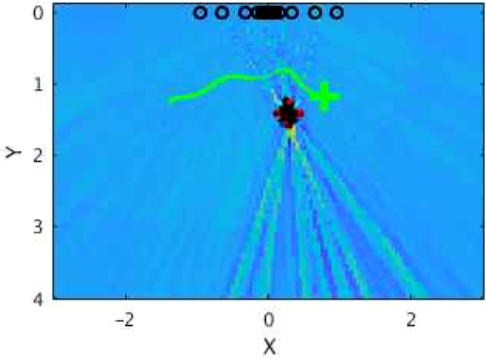}
\caption{Acoustic map (top view) at $t=1043$, on Task 3, recording 2, when a noise source is observed at the center of the room (black circles: microphones; black dots: particles; green line: annotated trajectory).}
\label{fig:centernoise}
\end{center}
\end{figure}

% Apart from the computational cost, sparsity of the pre-defined grid is another significant fact that might affect the performance. For example, in our experiment, the grid spacing is $2$ cm, where each point represents a $2$cm $\times 2$ cm $\times 2$ cm volume, thus leads to a $1.7$ $cm$ uncertainty in 3D. This uncertainty might be be reduced by using a denser grid, while compensating the computational cost. However, according to Table. \ref{tab:gridstep}, which shows the influence of the grid resolution on the localization accuracy and elapsed time, the localization accuracy doesn't have much variations when we increasing the grid resolution from $1cm$ to $10cm$, while the elapsed time shows an increasing in an exponential way.

\begin{table}[!tb]
\centering

\caption{Influence of the 3D grid resolution on the localization error (MAE in m) and the computational time (in minutes).}
\label{tab:gridstep}
\begin{tabular}{|c|c|c|c|c|c|c|c|}
\hline
\multicolumn{2}{|c|}{\textbf{resolution}  (m)} & \textbf{0.5} & \textbf{0.3} & \textbf{0.1} & \textbf{0.05} & \textbf{0.02} & \textbf{0.01} \\ \hline\hline
\multirow{2}{*}{T1}        &  time       & 0.18         & 0.17         & 0.18         & 0.30          & 2.24          & 28.83         \\ \cline{2-8} 
                               & error           & 0.83         & 0.43         & 0.24         & 0.18          & 0.19          & 0.20          \\ \hline\hline
\multirow{2}{*}{T3}        & time       & 0.13         & 0.13         & 0.17         & 0.60          & 8.49          & 135.89        \\ \cline{2-8} 
                               &  error           & 1.27         & 0.98         & 0.61         & 0.46          & 0.41          & 0.41          \\ \hline\hline
\end{tabular}
\end{table}

%%%%%%%%%%%%%%%%%%%%%%%%%%%%%%%%%
\section{Conclusion}
\label{sec:conclusion}

This document described our localization approach for the 2018 \ac{LOCATA} challenge. We discussed the importance of the choice of the window length, which depends on the speed of the sound source, and showed how localization accuracy varies with different microphone pairs. Moreover, to address the front-back ambiguity of a planar microphone array, we introduced an approach with a small computational overhead that is based on the temporal distribution of the \ac{GCF} peaks. This method is applicable to detect more than one turning frame resulting from the front-back ambiguity, by applying a sliding window, with different settings.

% -------------------------------------------------------------------------
% Either list references using the bibliography style file IEEEtran.bst
%\renewcommand*{\bibfont}{\footnotesize}
\bibliographystyle{IEEEtran}
\bibliography{refs18}
% %

% or list them by yourself
% \begin{thebibliography}{9}
%
% \bibitem{waspaa15web}
%   \url{http://www.waspaa.com}.
%
% \bibitem{IEEEPDFSpec}
%   {PDF} specification for {IEEE} {X}plore$^{\textregistered}$,
%   \url{http://www.ieee.org/portal/cms_docs/pubs/confstandards/pdfs/IEEE-PDF-SpecV401.pdf}.
%
% \bibitem{PDFOpenSourceTools}
%   Creating high resolution {PDF} files for book production with
%   open source tools,
%   \url{http://www.grassbook.org/neteler/highres_pdf.html}.
%
% \bibitem{eWilliams1999}
% E. Williams, \emph{Fourier Acoustics: Sound Radiation and Nearfield Acoustic
%   Holography}. London, UK: Academic Press, 1999.
%
% \bibitem{ieeecopyright}
%   \url{http://www.ieee.org/web/publications/rights/copyrightmain.html}.
%
% \bibitem{cJones2003}
% C. Jones, A. Smith, and E. Roberts, ``A sample paper in conference
%   proceedings,'' in \emph{Proc. IEEE ICASSP}, vol. II, 2003, pp. 803--806.
%
% \bibitem{aSmith2000}
% A. Smith, C. Jones, and E. Roberts, ``A sample paper in journals,''
%   \emph{IEEE Trans. Signal Process.}, vol. 62, pp. 291--294, Jan. 2000.
%
% \end{thebibliography}

\end{sloppy}
\end{document}

%% file: acronyms.tex
\newacro{TDOA}[TDoA]{Time Difference of Arrival}
\newacro{SRP}[SRP]{Steered Response Power}
\newacro{MUSIC}[MUSIC]{MUltiple SIgnal Classification}
\newacro{GCC}[GCC]{Generalized Cross Correlation}
\newacro{GCC-PHAT}[GCC-PHAT]{Generalized Cross Correlation with Phase Transform}
\newacro{MAE}[MAE]{Mean Absolute Error}
\newacro{PF}[PF]{Particle Filter}
\newacro{PHAT}[PHAT]{PHAse Transform}
\newacro{CM}[CM]{Coherence Measure}
\newacro{LMS}[LMS]{Least Mean Square}
\newacro{DoA}[DoA]{Direction of Arrival}
\newacro{GCF}[GCF]{Global Coherence Field}
\newacro{DICIT}[DICIT]{Distant-talking Interfaces for Control of Interactive TV}
\newacro{LOCATA}[LOCATA]{LOCalization And TrAcking}
\newacro{STFT}[STFT]{Short-Time-Fourier-Transform}
\newacro{DFT}[DFT]{Discrete Fourier Transform}
\newacro{KF}[KF]{Kalman Filter}
\newacro{MUSIC}[MUSIC]{MUltiple SIgnal Classification}
\newacro{SIR}[SIR]{Sequential Importance Re-sampling}
\newacro{SSL}[SSL]{Sound Source Localization}

%% file: latex18.bbl
\begin{thebibliography}{10}
\providecommand{\url}[1]{#1}
\def\UrlFont{\rmfamily}
\providecommand{\newblock}{\relax}
\providecommand{\bibinfo}[2]{#2}
\providecommand\BIBentrySTDinterwordspacing{\spaceskip=0pt\relax}
\providecommand\BIBentryALTinterwordstretchfactor{4}
\providecommand\BIBentryALTinterwordspacing{\spaceskip=\fontdimen2\font plus
\BIBentryALTinterwordstretchfactor\fontdimen3\font minus
  \fontdimen4\font\relax}
\providecommand\BIBforeignlanguage[2]{{%
\expandafter\ifx\csname l@#1\endcsname\relax
\typeout{** WARNING: IEEEtran.bst: No hyphenation pattern has been}%
\typeout{** loaded for the language `#1'. Using the pattern for}%
\typeout{** the default language instead.}%
\else
\language=\csname l@#1\endcsname
\fi
#2}}

\bibitem{locata2018heinrich}
H.~W. Lollmann, C.~Evers, A.~Schmidt, H.~Mellmann, H.~Barfuss, P.~A. Naylor,
  and W.~Kellermann, ``The {LOCATA} challenge data corpus for acoustic source
  localization and tracking,'' in \emph{IEEE Sensor Array and Multichannel
  Signal Processing Workshop (SAM)}, Sheffield, UK, July 2018.

\bibitem{Veaux2018english}
\BIBentryALTinterwordspacing
C.~Veaux, J.~Yamagishi, and K.~MacDonald. (2018) English multi-speaker corpus
  for {CSTR} voice cloning toolkit. [Online]. Available:
  \url{http://homepages.inf.ed.ac.uk/jyamagis/page3/page58/page58.html}
\BIBentrySTDinterwordspacing

\bibitem{Ottoy2016an}
G.~Ottoy and L.~De~Strycker, ``An improved {2D} triangulation algorithm for use
  with linear arrays,'' \emph{IEEE Sensors Journal}, vol.~16, no.~23, pp.
  8238--8243, December 2016.

\bibitem{omologo1998spoken}
M.~Omologo, P.~Svaizer, and R.~De~Mori, ``Acoustic transduction,'' in
  \emph{Spoken Dialogue with Computer}.\hskip 1em plus 0.5em minus 0.4em\relax
  Academic Press, 1998, ch.~2, pp. 1--46.

\bibitem{knapp1976generalized}
C.~Knapp and G.~Carter, ``The generalized correlation method for estimation of
  time delay,'' \emph{IEEE Transactions on Acoustics, Speech, and Signal
  Processing}, vol.~24, no.~4, pp. 320--327, 1976.

\bibitem{omologo1994acoustic}
M.~Omologo and P.~Svaizer, ``Acoustic event localization using a
  crosspower-spectrum phase based technique,'' in \emph{Proc. of {IEEE} Int.
  Conf. on Audio, Speech and Signal Processing}, Adelaide, SA, Australia, April
  1994, pp. 273--276.

\bibitem{arulampalam2002tutorial}
M.~S. Arulampalam, S.~Maskell, N.~Gordon, and T.~Clapp, ``A tutorial on
  particle filters for online nonlinear/non-{G}aussian {B}ayesian tracking,''
  \emph{IEEE Transactions on Signal Processing}, vol.~50, no.~2, pp. 174--188,
  2002.

\bibitem{qian20173d}
X.~Qian, A.~Brutti, M.~Omologo, and A.~Cavallaro, ``{3D} audio-visual speaker
  tracking with an adaptive particle filter,'' in \emph{Proc. of {IEEE} Int.
  Conf. on Audio, Speech and Signal Processing}, New Orleans, LA, USA, March
  2017.

\bibitem{brutti2010WOZ}
A.~Brutti, L.~Cristoforetti, W.~Kellermann, L.~Marquardt, and M.~Omologo,
  ``{WOZ} acoustic data collection for interactive {TV},'' \emph{Language
  Resources and Evaluation}, vol.~44, no.~3, pp. 205--219, September 2010.

\bibitem{trees2002optimum}
H.~L. Van~Trees, \emph{Optimum Array Processing: Part IV of Detection,
  Estimation, and Modulation Theory}, Wiley, New York, USA, 2002.

\bibitem{dmochowski2007broadband}
J.~P. Dmochowski, J.~Benesty, and S.~Affes, ``Broadband {MUSIC}: Opportunities
  and challenges for multiple source localization,'' in \emph{Proc. of IEEE
  Workshop on Applications of Signal Processing to Audio and Acoustics
  (WASPAA)}, New Paltz (New York), USA, October 2007, pp. 18--21.

\end{thebibliography}
